\documentclass[aip,jap,reprint]{revtex4-1}

\usepackage{graphicx}
\begin{document}

\title{Crystal structure, local structure, and defect structure of Pr-doped SrTiO$_3$}



\author{I. A. Sluchinskaya}
\email[]{E-mail: irinasluch@nm.ru}
\author{A. I. Lebedev}
\affiliation{Moscow State University, Physics Department, Leninskie gory, 119991 Moscow, Russia}
\author{A. Erko}
\affiliation{Helmholtz-Zentrum Berlin, Albert-Einstein-Str. 15, 12489 Berlin, Germany}


\date{\today}

\begin{abstract}
X-ray diffraction studies showed that the structure of
(Sr$_{1-x}$Pr$_x$)TiO$_3$ solid solutions at 300~K changes from the cubic $Pm3m$
to the tetragonal $I4/mcm$ with increasing $x$. The analysis of XANES and EXAFS
spectra of the solid solutions revealed that Pr ions are predominantly in the 3+
oxidation state, they substitute for Sr atoms and are on-center regardless of
the preparation conditions. The weak dependence of the lattice parameter in
(Sr$_{1-x}$Pr$_x$)TiO$_3$ on the Pr concentration was explained by the competition
between the relaxation of the Sr--O bond length, which results from the difference
in ionic radii of Sr and Pr ions, and the repulsion of positively charged Pr$^{3+}$
and Ti$^{4+}$ ions. It was shown that the most important defects
in the crystals are charged Sr vacancies and SrO planar faults;
praseodymium does not enter the Sr sites in the planar faults.
\end{abstract}


\maketitle 


\section{\label{sec1}Introduction}

Strontium titanate SrTiO$_3$ is a perovskite-type crystal widely used in
electronics and optoelectronics. Because of its high room-temperature
dielectric constant ($\sim$300) that strongly depends on the electric field,
strontium titanate is used as a high-$k$ gate dielectric in field-effect
transistors and electric-field-tunable capacitors working at microwaves.

Interesting luminescence properties of rare-earth-doped strontium titanate
have long been a subject of intensive studies. In particular, doping SrTiO$_3$
with Pr is known to result in the red luminescence. Although the luminescence of
SrTiO$_3$(Pr) itself is rather weak, co-doping the samples with oxides of the
group-III elements (Al, B, Ga, In) enhances the luminescence efficiency up to
200~times.~\cite{JApplPhys.86.5594,ApplPhysLett.78.655}  Red-emitting
SrTiO$_3$(Pr$^{3+}$,Al) phosphors are now promising materials for field-emission
and vacuum fluorescent displays. However, the
influence of co-doping on luminescent properties and
the nature of defects acting as non-radiative centers in Pr-doped SrTiO$_3$
are not well understood.
The mechanisms responsible for maintaining charge neutrality in this material
have not been identified yet.

Strontium titanate is a model system for studying physical phenomena resulting
in ferroelectricity. It is an incipient ferroelectric with a cubic $Pm3m$
structure at 300~K, in which quantum fluctuations stabilize the paraelectric
phase up to the lowest temperatures. At about 105~K SrTiO$_3$ undergoes
the structural (antiferrodistortive) $Pm3m \to I4/mcm$ phase transition
associated with the oxygen octahedra rotations.~\cite{LB}
This transition is, however, of nonpolar character and weakly influences the
dielectric properties. Ferroelectricity in SrTiO$_3$ can be induced by the
application of external electric~\cite{JPhysCondensMatter.8.4673} or strain
fields,~\cite{PhysRevB.13.271}  doping with
impurities,~\cite{PhysRev.124.1354,PhysRevLett.52.2289,PhysRevB.54.3151}
or isotope substitution.~\cite{PhysRevLett.82.3540}  Due to the large lattice
polarizability, SrTiO$_3$ is very sensitive to doping. The substitution of
Sr$^{2+}$ with Pb$^{2+}$, Ba$^{2+}$, Ca$^{2+}$, Cd$^{2+}$, Mn$^{2+}$, and
Bi$^{3+}$ ions strongly influences the dielectric properties of SrTiO$_3$
and can induce various types of polar phases (dipole glass,
ferroelectric).~\cite{Lemanov.NATO1999.329,PhysSolidState.43.2146,
PhysSolidState.46.1442,JApplPhys.98.056102,PhysRevB.59.6670}  In most cases,
however, the temperature of the ferroelectric phase transition in doped
SrTiO$_3$ is lower than 300~K.

It was recently reported that the ferroelectric phase transition temperature
in Pr-doped SrTiO$_3$ can exceed 300~K.~\cite{JApplPhys.97.104109}  In
(Sr$_{1-x}$Pr$_x$)TiO$_3$ samples with $x = {}$0.025--0.075 Duran \emph{et al.}
observed a dielectric anomaly at about 515~K and dielectric hysteresis loops
at room temperature, which were attributed to the ferroelectric phase transition.
The crystal structure of the samples, however, remained cubic at 300~K, and the
temperature of the dielectric anomaly changed by only 10~K when the Pr
concentration was increased by three times. This influence of Pr doping differs
significantly from the influence of other impurities on the temperature of the
ferroelectric phase transition in doped SrTiO$_3$.~\cite{Lemanov.NATO1999.329,
PhysSolidState.43.2146,PhysSolidState.46.1442,JApplPhys.98.056102}  According
to the XPS spectra,~\cite{JApplPhys.97.104109} Pr atoms in SrTiO$_3$ can be in two
oxidation states (3+ and 4+). The subsequent synchrotron radiation x-ray
diffraction study of a (Sr$_{1-x}$Pr$_x$)TiO$_3$ sample with higher Pr
concentration ($x = 0.15$) revealed a tetragonal lattice distortion, which
was attributed to the appearance of the $P4mm$ phase.~\cite{JPhysCondensMatter.20.085219}

X-ray and neutron powder diffraction studies of the samples with $x \ge 0.09$
by another experimental group~\cite{PhysRevB.78.092102,PhysRevB.79.144122}
revealed superstructure peaks in the diffraction patterns, and the structure of
the samples was determined to be $I4/mcm$. It was concluded that doping
SrTiO$_3$ with Pr increases the temperature of the phase transition to the
$I4/mcm$ phase, and at $x = 0.05$ it is close to 300~K.~\cite{PhysRevB.78.092102}
Raman studies of these
samples~\cite{PhysRevB.76.224109,PhysRevB.78.092102} revealed weak $E_g$ mode,
which originates from the $T_{2u}$ silent mode of the $Pm3m$ phase and becomes Raman
active in the $I4/mcm$ phase. Monotonic change of the frequency of the (formally
forbidden in the $Pm3m$ phase) soft TO$_1$ mode in Raman spectra of
(Sr$_{0.975}$Pr$_{0.025}$)TiO$_3$ without any features at the temperature
of the dielectric anomaly~\cite{PhysRevB.76.224109}  proves the absence of the
displacive ferroelectric phase transition in SrTiO$_3$(Pr). Observation of
the dielectric relaxation~\cite{PhysRevB.76.224109}  showed that the system
can be regarded as a high-temperature relaxor ferroelectric with the
temperature of transition to the relaxor state independent of the
structural phase transition temperature.~\cite{PhysRevB.79.144122}

Dielectric studies~\cite{JApplPhys.107.094108} of (Sr$_{1-x}$Pr$_x$)TiO$_3$
samples with $0 \le x \le 0.03$ revealed a ``colossal'' room-temperature
dielectric constant of about 3000 at 1~KHz in the sample with $x = 0.01$. The
results of this work, however, differed considerably from those of previous
studies:~\cite{JApplPhys.97.104109,JPhysCondensMatter.20.085219,PhysRevB.76.224109,
PhysRevB.78.092102, PhysRevB.79.144122}  instead of peaks in the real part of the
complex dielectric constant
$\varepsilon = \varepsilon^\prime + i\varepsilon^{\prime\prime}$, a monotonic
increase of $\varepsilon^\prime$ with noticeable inflections accompanied by the
corresponding peaks in $\varepsilon^{\prime\prime}$ were observed
in the 0--500$^\circ$C temperature range.

To explain the observed dielectric anomaly, several models have been
proposed: (1)~displacive ferroelectric phase transition in the
bulk;~\cite{JApplPhys.97.104109}  (2)~appearance of polar nanoregions around
various defects such as Pr$^{3+}$/Pr$^{4+}$--$V_{\rm Sr}$ complexes, off-center
Pr$_{\rm Sr}$ ions, and Ti$^{4+}$ ions displaced as a result of the Sr$\to$Pr
substitution;~\cite{PhysRevB.76.224109}  (3)~capacitive effect arising from
the semiconducting nature of grains and insulating grain boundaries in ceramic
samples.~\cite{JApplPhys.107.094108}

In this paper, the crystal structure of Pr-doped SrTiO$_3$, the structural
position and the oxidation state of Pr ions, and the mechanisms responsible for
maintaining charge neutrality upon heterovalent substitution are studied using
x-ray diffraction, extended x-ray absorption fine structure (EXAFS), and x-ray
absorption near-edge structure (XANES) techniques. A model of the defect
structure which explains main properties of these samples is proposed, and
possible interpretations of the dielectric anomaly are critically analyzed.

\section{Experimental}

Samples with a nominal composition of (Sr$_{1-x}$Pr$_x$)TiO$_3$ ($x = {}$0.05,
0.15, and 0.3) and Sr(Ti$_{1-x}$Pr$_x$)O$_3$ ($x = 0.05$) were prepared by the
solid-state reaction method from SrCO$_3$, Pr$_6$O$_{11}$, and nanocrystalline
TiO$_2$ which was synthesized by hydrolysis of tetrapropylorthotitanate and
dried at 500$^\circ$C. Starting materials were weighed in necessary proportions,
mixed, ground under acetone, and calcined in air at 1100$^\circ$C for 8~h. The
calcined powders were ground once more and annealed again at
1100--1600$^\circ$C. The annealing times were 8~h for 1100$^\circ$C,
4~h for 1300 and 1400$^\circ$C, 2~h for 1600$^\circ$C. The temperature
of 1100$^\circ$C was found insufficient to enter the praseodymium into reaction;
solid solutions were formed starting from 1300$^\circ$C.

EXAFS and XANES spectra were obtained at KMC-2 station of the BESSY
synchrotron radiation source at the Pr $L_{\rm II}$-edge (6440~eV) and the
Ti $K$-edge (4966~eV) at 300~K. Spectra were collected in fluorescence mode
using an energy-dispersive R{\"o}ntec X-flash detector with 10~mm$^2$ active
area. The choice of the Pr $L_{\rm II}$-edge instead of the more common
$L_{\rm III}$-edge was motivated by the following. The energies of the
Pr $L_{\alpha 1}$ fluorescence line (exited at the $L_{\rm III}$-edge) and
of the Ti $K_{\beta}$ fluorescence line are very close. This strongly
complicates the measurements. Selecting of the Pr
$L_{\rm II}$-edge results in excitation of the $L_{\beta 1}$ fluorescence line,
whose energy differs by 560~eV from the energy of the Ti $K_{\beta}$ fluorescence
line. This is sufficient to separate the signals from Pr and Ti atoms. The
spectra at the Pr $L_{\rm II}$-edge were recorded in a limited
energy range (6400--6840~eV) because of the proximity of the Pr $L_{\rm I}$-edge.
EXAFS spectra were processed in two ways. First, they were analyzed in our
traditional way~\cite{PhysRevB.55.14770} using FEFF6 software~\cite{FEFF} for
calculating the scattering amplitudes and phases. Second, the data were
pre-processed using ATHENA software and were fitted using ARTEMIS
software~\cite{IFEFFIT} to the theoretical curves computed for a given structural
model; in this approach, the scattering amplitudes and phases were also
calculated using FEFF6. The results obtained by both methods agreed well.

\section{Results}

X-ray diffraction studies showed that Sr(Ti$_{0.95}$Pr$_{0.05}$)O$_3$ samples
annealed at 1300--1600$^\circ$C were not single-phase; additional peaks of
the Sr$_3$Ti$_2$O$_7$ Ruddlesden-Popper phase and Pr$_6$O$_{11}$ were observed
in their diffraction patterns. (Sr$_{1-x}$Pr$_x$)TiO$_3$ samples with
$x = {}$0.05--0.15 were single-phase, whereas the sample with $x = 0.3$
contained a small amount of PrO$_{2-\delta}$ and SrPr$_4$Ti$_5$O$_{17}$
second phases.

\begin{figure}
\centering
\includegraphics{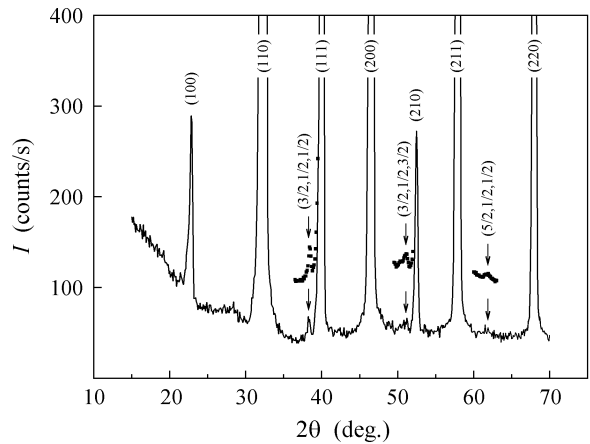}
\caption{\label{fig1}X-ray diffraction pattern of (Sr$_{0.85}$Pr$_{0.15}$)TiO$_3$
sample annealed at 1400$^\circ$C. Fragments of the diffraction pattern recorded
with the integration time of 300~s are shifted along the vertical axis and
shown by squares. Calculated positions of the superstructure peaks are denoted
by arrows.}
\end{figure}

Along with peaks characteristic of the cubic perovskite structure, additional
superstructure reflections, whose intensity increased with increasing Pr
concentration, were observed in the diffraction patterns of
(Sr$_{1-x}$Pr$_x$)TiO$_3$ samples with $x = {}$0.15--0.3. These peaks are
denoted by arrows in Fig.~\ref{fig1} and can be indexed on the cubic lattice
as (3/2,~1/2,~1/2), (3/2,~1/2,~3/2), and (5/2,~1/2,~1/2) reflections. Following
Glazer,~\cite{ActaCrystA.31.756} the appearance of these
superstructure reflections indicates the rotation of the oxygen octahedra around
the $c$ axis according to the $a^0a^0c^-$ tilt system, which results in a lowering
of symmetry from cubic $Pm3m$ to tetragonal $I4/mcm$ and in a doubling of the
unit cell. Since the structural phase transition to the same phase
occurs in undoped SrTiO$_3$ at 105~K, one can conclude that doping SrTiO$_3$
with Pr results in the increase of the phase transition temperature, and this
temperature exceeds 300~K in samples with $x \ge 0.15$. This conclusion agrees
with that drawn from the neutron diffraction study,~\cite{PhysRevB.79.144122}
but it can be noted that in this work a full set of superstructure reflections
was observed in x-ray diffraction.

The lattice parameters of (Sr$_{1-x}$Pr$_x$)TiO$_3$ samples annealed at
1400$^\circ$C were $a = 3.897$~{\AA} for $x = 0.05$, 3.898~{\AA} for $x = 0.15$,
and 3.894~{\AA} for $x = 0.3$.%
    \footnote{The $a$ and $c$ parameters of the tetragonal phase were so close
    that they could not be determined independently. Therefore, we considered
    the unit cell as a pseudocubic one.}
It is seen that the lattice parameter decreases slowly with increasing Pr
concentration. According to
Refs.~\onlinecite{JApplPhys.97.104109,PhysRevB.78.092102,PhysRevB.79.144122},
the lattice parameter of (Sr$_{1-x}$Pr$_x$)TiO$_3$ is nearly independent of $x$.

\begin{figure}
\centering
\includegraphics{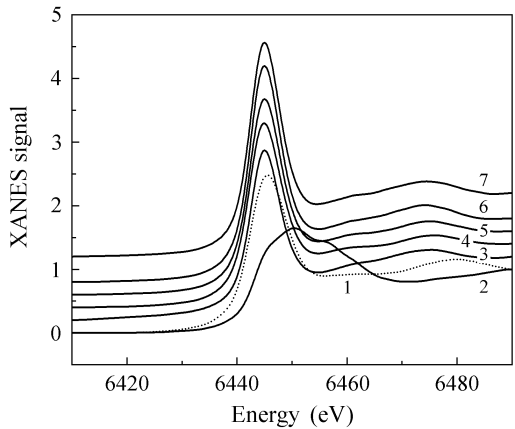}
\caption{\label{fig2}XANES spectra of SrTiO$_3$(Pr) samples and reference
compounds of tri- and tetra-valent praseodymium recorded at the Pr
$L_{\rm II}$-edge at 300~K. 1~--- Pr$_2$Ti$_2$O$_7$, 2~--- BaPrO$_3$, 3, 6, 7~---
(Sr$_{0.95}$Pr$_{0.05}$)TiO$_3$ samples annealed at 1600, 1400, and 1300$^\circ$C,
respectively, 4~--- (Sr$_{0.7}$Pr$_{0.3}$)TiO$_3$ sample annealed at 1400$^\circ$C,
5~--- (Sr$_{0.85}$Pr$_{0.15}$)TiO$_3$ sample annealed at 1400$^\circ$C.}
\end{figure}

To determine the oxidation state of Pr ions in our samples, their XANES
spectra were compared with those of the reference compounds. The spectra for
five (Sr$_{1-x}$Pr$_x$)TiO$_3$ samples and Pr$^{3+}$ and Pr$^{4+}$ reference
compounds (Pr$_2$Ti$_2$O$_7$ and BaPrO$_3$, respectively)
are shown in Fig.~\ref{fig2}. Comparison of these spectra shows that the
oxidation state of the Pr ion in our samples is predominantly 3+ and does not
depend on the Pr concentration and the preparation conditions. This differs from the
behavior of the Mn impurity in SrTiO$_3$, whose oxidation state and structural
position strongly depended on the preparation
conditions.~\cite{JETPLett.89.457,BullRASPhys.74.1235}

\begin{table*}
\caption{\label{table1}Structural parameters obtained from the EXAFS data analysis
($R_i$ is the distance to the $i$-th shell, $\sigma_i^2$ is the Debye-Waller factor
for this shell) and mean interatomic distances in the perovskite structure with
the lattice parameters taken from x-ray measurements. The $R$-factor indicates
the goodness of fit of the theoretical curves to the Fourier-filtered spectra.}
\begin{ruledtabular}
\begin{tabular}{cccccccc}
Shell  & \multicolumn{2}{c}{(Sr$_{0.95}$Pr$_{0.05}$)TiO$_3$} & \multicolumn{2}{c}{(Sr$_{0.95}$Pr$_{0.05}$)TiO$_3$} & \multicolumn{2}{c}{(Sr$_{0.85}$Pr$_{0.15}$)TiO$_3$} & (Sr$_{0.95}$Pr$_{0.05}$)TiO$_3$, \\
       & \multicolumn{2}{c}{annealed at 1300$^\circ$C} & \multicolumn{2}{c}{annealed at 1600$^\circ$C} & \multicolumn{2}{c}{annealed at 1400$^\circ$C} & x-ray data \\
       & $R_i$ ({\AA}) & $\sigma_i^2$ ({\AA}$^2$) & $R_i$ ({\AA}) & $\sigma_i^2$ ({\AA}$^2$) & $R_i$ ({\AA}) & $\sigma_i^2$ ({\AA}$^2$) & $R_i$ ({\AA}) \\
\hline
Pr--O  & 2.629(9) & 0.017(1) & 2.623(7) & 0.014(1) & 2.608(15)& 0.020(2) & 2.756 \\
Pr--Ti & 3.394(5) & 0.003(1) & 3.388(4) & 0.003(1) & 3.404(8) & 0.004(1) & 3.375 \\
Pr--Sr & 3.887(34)& 0.022(5) & 3.926(13)& 0.012(2) & 3.930(38)& 0.018(5) & 3.897 \\
$R$-factor & \multicolumn{2}{c}{0.0041} & \multicolumn{2}{c}{0.0037} & \multicolumn{2}{c}{0.0112} & --- \\
\end{tabular}
\end{ruledtabular}
\end{table*}

\begin{figure}
\centering
\includegraphics{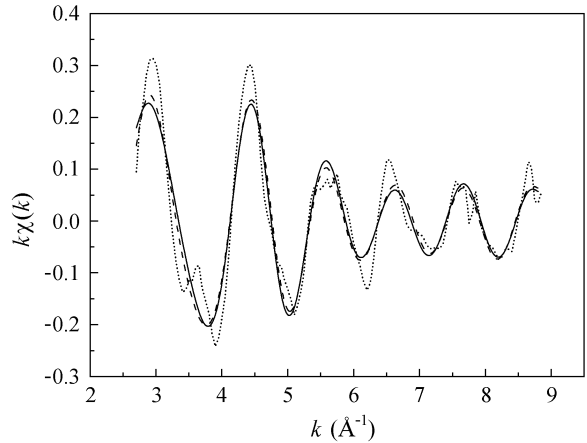}
\caption{\label{fig3}EXAFS spectra obtained at the Pr $L_{\rm II}$-edge at
300~K for (Sr$_{0.85}$Pr$_{0.15}$)TiO$_3$ sample annealed at 1400$^\circ$C.
The dashed line shows the Fourier-filtered spectrum ($R = {}$1.3--4.0~{\AA},
three nearest shells) and the solid line is its best theoretical fit. The raw
data are shown by the dotted line.}
\end{figure}

To determine the structural position of the Pr impurity, EXAFS spectra were
analyzed. Typical Fourier-filtered EXAFS spectrum as a function of the
photoelectron wave vector $k$ for the (Sr$_{0.85}$Pr$_{0.15}$)TiO$_3$ sample and
its best theoretical fit are shown in Fig.~\ref{fig3}. The best
agreement between the experimental and calculated data was obtained for a model
in which the Pr atoms substitute for the Sr atoms. Interatomic distances and
Debye-Waller factors for three nearest shells are given in Table~\ref{table1}.
Comparison of the obtained interatomic distances with those calculated for an
ideal perovskite structure with the lattice parameter taken from x-ray
measurements shows that a strong relaxation around the impurity atom occurs only
in the first shell (the average relaxation is $\Delta R \approx -0.136$~{\AA}).
In contrast, a slight increase in the average interatomic distance
($\Delta R \approx +0.02$~{\AA}) is found in the second shell. In the third shell,
the accuracy of determination of the structural parameters is not sufficient to draw
any conclusions; however, the interatomic distance for this shell is consistent
with x-ray data (Table~\ref{table1}).

Small values of the Debye-Waller factors for the second shell (Table~\ref{table1}),
which are typical for thermal vibrations in perovskites at 300~K, enable to
completely exclude the off-centering of the Pr atoms. Unexpectedly
large Debye-Waller factor for the first shell in the sample with $x = 0.15$
can be explained by static distortions of the Pr--O bond lengths resulting from
the rotation of the oxygen octahedra in the $I4/mcm$ phase. In the sample with
$x = 0.05$, high Debye-Waller factor may indicate large thermal fluctuations
of the rotation angle because at this Pr concentration, the temperature of the
structural phase transition is close to 300~K.~\cite{PhysRevB.78.092102}

To check for the possibility of the Pr incorporation into the $B$~sites of the
perovskite structure, experimental EXAFS spectra were compared with the spectra
calculated in the model in which the impurity enters both the $A$ and $B$~sites
simultaneously. It turned out that with increasing concentration of Pr at the $B$~sites,
the agreement of the curves becomes worse. This means that the incorporation
of Pr into the $B$~sites is very unlikely, apparently because of the strong
difference in ionic radii of Ti$^{4+}$ (0.605~{\AA}) and Pr$^{3+}$ (0.99~{\AA})
in octahedral coordination.~\cite{ActaCrystA.32.751}

\section{Discussion}

\begin{figure}
\centering
\includegraphics{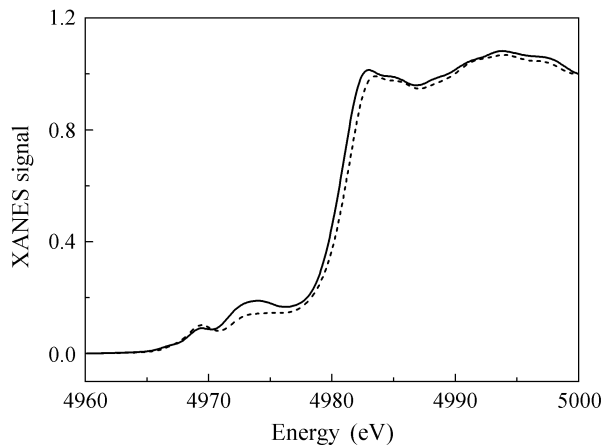}
\caption{\label{fig4}XANES spectra recorded at the Ti $K$-edge at 300~K for
(Sr$_{0.85}$Pr$_{0.15}$)TiO$_3$ sample annealed at 1400$^\circ$C (full line) and
undoped SrTiO$_3$ sample (dashed line).}
\end{figure}

X-ray studies performed in this work confirmed previous
results~\cite{PhysRevB.78.092102,PhysRevB.79.144122} that doping SrTiO$_3$
with Pr increases the temperature of the structural phase transition from the cubic
$Pm3m$ to the tetragonal $I4/mcm$ phase.

The weak effect of Pr doping on the lattice parameter of (Sr$_{1-x}$Pr$_x$)TiO$_3$
is an unusual feature of these samples.~\cite{JApplPhys.97.104109,
PhysRevB.78.092102,PhysRevB.79.144122}  In our x-ray measurements, the decrease
of the lattice parameter was less than 0.008~{\AA} in the whole solubility range
($a = 3.905$~{\AA} in bulk SrTiO$_3$).

We compare the interatomic distances obtained from x-ray and EXAFS measurements. The
average difference between the x-ray and EXAFS Pr--O distances in the first
shell ($-$0.136~{\AA}) is very close to the difference between ionic radii of
Pr$^{3+}$ and Sr$^{2+}$. Because of the lack of data on the ionic radius of the
Pr$^{3+}$ ion in 12-fold coordination, we used the data for 8-fold coordination
for both ions, $R_{\rm Sr^{2+}} = 1.26$~{\AA} and
$R_{\rm Pr^{3+}} = 1.126$~{\AA}.~\cite{ActaCrystA.32.751}  The difference of
these ionic radii, 0.134~{\AA}, coincides with the relaxation in the first
shell obtained from our EXAFS measurements.

At the same time, the EXAFS interatomic distances in the second shell
turned out a little bit larger than the mean interatomic distances in
the perovskite structure. This can be explained as follows. According to
the second Pauling's rule,~\cite{JAmChemSoc.51.1010} the perovskite structure
is electrostatically compensated, which means that the sum of the strengths of
the electrostatic valence bonds of Sr$^{2+}$ and Ti$^{4+}$ cations which reach
the O$^{2-}$ anion is equal to the ionic charge of the anion. Substitution of
Sr$^{2+}$ by Pr$^{3+}$ disturbs the local charge compensation and results in the
repulsion of positively charged Pr$^{3+}$ and Ti$^{4+}$ ions. So, the Pr--Ti
interatomic distance increases. The fact that we did not reveal the
incorporation of Pr atoms into the \emph{B}~sites of the perovskite structure
excludes the explanation~\cite{PhysRevB.78.092102}  that the weak dependence of
the lattice parameter in (Sr$_{1-x}$Pr$_x$)TiO$_3$ on $x$ is caused by the
partial substitution of Ti$^{4+}$ ions by larger Pr$^{3+}$ ions. In our opinion,
this weak dependence is a consequence of
two competing effects, the relaxation in the first shell resulting form
the difference in ionic radii, and the repulsion of positively charged Pr$^{3+}$
and Ti$^{4+}$ ions.

As the main oxidation state of the Pr ions in our samples is 3+, this implies
the existence of some charged defects to maintain charge neutrality of the
sample. The vacancies $V_{\rm Sr}$, $V_{\rm Ti}$, and $V_{\rm O}$ can act as
such defects, as well as Ti atoms in the 3+ oxidation state.

The possibility of reduction of the Ti ion to Ti$^{3+}$ state can be excluded
because otherwise the samples would darken (our samples are light yellow).
To clarify the role of the oxygen vacancies, the pre-edge structure at the
Ti $K$-edge in the (Sr$_{0.95}$Pr$_{0.05}$)TiO$_3$ sample was studied. The
experiment did not reveal any increase in the intensity of the $1s \to 3d$
($e_g$) dipole-forbidden
transition at 4969~eV (Ref.~\onlinecite{PhysSolidState.51.991})
(Fig.~\ref{fig4}), thus indicating that the symmetry of the oxygen octahedra
was preserved (the symmetry would be broken by the presence of $V_{\rm O}$
vacancies). An increase in the intensity of two unresolved peaks
at 4972--4975~eV, which are attributed to the dipole-forbidden
transitions of the Ti $1s$ electron into the $t_{2g}$ and $e_g$ orbitals of
neighboring TiO$_6$ octahedra,~\cite{JPhysCondensMatter.10.9561}
is probably a consequence of the
Pr-induced disorder in our samples. The formation of Ti vacancies in our
samples is improbable because otherwise the excess of Ti would precipitate as
the TiO$_2$ phase; the peaks of this phase were absent in our x-ray diffraction patterns.
Therefore, the most important charged defects that maintain charge neutrality
in Pr-doped SrTiO$_3$ are Sr vacancies.

The existence of Sr vacancies in (Sr$_{1-x}$Pr$_x$)TiO$_3$ samples with high
$x$ implies the precipitation of excess Sr. For example, in the sample with
$x = 0.15$, the concentration of $V_{\rm Sr}$ should be equal to 7.5\%, so that
this amount of excess strontium should be precipitated as an oxide phase
which can be easily detected in x-ray experiments. However, no additional peaks
were observed in x-ray diffraction patterns of the
(Sr$_{0.85}$Pr$_{0.15}$)TiO$_3$ sample. At the same time, it is well known
that in the SrTiO$_3$--SrO system the excess Sr can form so-called planar
faults, which are stacking faults consisting of one extra SrO layer
inserted between SrO and TiO$_2$ layers in the SrTiO$_3$ structure and intergrown
coherently with the adjacent layers.~\cite{PhilMagA.75.833,JMaterRes.15.2131}
In the case of a regular arrangement of these extra layers, the structures called
the Ruddlesden-Popper phases with a general formula Sr$_{n+1}$Ti$_n$O$_{3n+1}$
are formed.~\cite{ActaCryst.11.54}  The planar faults were observed using
electron microscopy even in stoichiometric
SrTiO$_3$.~\cite{JSolidStateChem.21.293}  In this work, the
Ruddlesden-Popper phases were observed in x-ray diffraction only in
Sr(Ti$_{1-x}$Pr$_x$)O$_3$ samples annealed at temperatures insufficient for
dissolving praseodymium. In single-phase (Sr$_{1-x}$Pr$_x$)TiO$_3$ samples, we
believe that excess Sr forms planar faults as well. But in this case, the
absence of the diffraction peaks of the Ruddlesden-Popper phase simply indicates the
absence of the long-range order in the arrangement of SrO planar faults.
We think that randomly distributed Pr$^{3+}$--$V_{\rm Sr}$ complexes
act as pinning centers for the motion of SrO planar faults, and so their regular
arrangement cannot be achieved.

From the previous discussion it follows that there are two possible positions
of Sr atoms in our samples: the Sr sites in the SrTiO$_3$ perovskite layers and
the Sr sites in the SrO planar faults. In the Ruddlesden-Popper phases, different impurity
atoms substitute for different Sr sites: for example, Ba atoms predominantly enter
the SrTiO$_3$ layers, thus forming a solid solution,~\cite{JMaterRes.24.2596}
whereas Ca atoms predominantly enter the SrO planar
faults.~\cite{JAmCeramSoc.81.33,JMaterRes.24.2596}  To refine the position of the
Pr atoms in our heavily doped samples, we additionally analyzed our EXAFS
spectra in the model in which Pr atoms substitute for Sr atoms in the SrO planar
faults. Initial structural positions of atoms were taken from the crystal
structure of the Sr$_3$Ti$_2$O$_7$ Ruddlesden-Popper phase. Because of the similarity
between the crystal structures of SrTiO$_3$ and the Ruddlesden-Popper phase (a half of
the Ruddlesden-Popper-phase structure nearly coincides with the perovskite structure),
the calculated EXAFS spectra for these structures
were similar. Nevertheless, a better agreement with the experimental EXAFS spectra
was obtained for the model in which the Pr atoms substitute for the Sr atoms in the SrTiO$_3$ layers.
This is not surprising because to incorporate the Pr atoms into the Sr sites in the SrO planar
faults, these planar faults should be first created by the Sr$\to$Pr substitution
in bulk SrTiO$_3$.

The defect model proposed in this work explains the influence of
addition of Al and other trivalent impurities (Ga, In, B) on the intensity of the
red luminescence in Pr-doped SrTiO$_3$. As was mentioned in Sec.~\ref{sec1},
co-doping SrTiO$_3$(Pr) with the trivalent impurities can enhance the luminescence
efficiency up to 200~times. We think that both Sr vacancies and SrO planar faults
act as non-radiative centers in SrTiO$_3$(Pr). The addition of trivalent impurities
results in two effects: (1)~partial substitution of titanium by trivalent
impurity maintains charge neutrality of the samples and excludes the
creation of Sr vacancies, and (2)~x-ray amorphous phase based on the oxide of
trivalent impurity ``absorbs'' the planar faults. As a result, the addition of
trivalent impurity ``purifies'' the grains of SrTiO$_3$ from point and planar
defects, thus creating optimal conditions for radiative recombination.

An increase of the structural phase transition temperature $T_s$ when doping
SrTiO$_3$ with Pr is not surprising because the ionic radius of Pr$^{3+}$ is
less than that of Sr$^{2+}$: the decrease of the $A$~atom size in the perovskite
structure decreases the tolerance factor, and the instability of the structure
against octahedra rotations increases. More surprising is the fact that the
$dT_s/dx$ rate for Pr doping is higher than that for Ca doping (the impurities have the same
ionic radius). For Pr doping, the temperature $T_s$ reaches 300~K at
$x = 0.05$,~\cite{PhysRevB.78.092102} whereas for Ca doping $T_s$ reaches 300~K
only at $x = 0.1$.~\cite{PhysRev.124.1354}  We believe that such a strong
effect of Pr doping on $T_s$ is due to the presence of Sr vacancies which
influence the octahedra rotation in a manner similar to the influence of small \emph{A}
cations.

In this work, we considered exclusively the structural properties of Pr-doped
SrTiO$_3$ and established that Pr is not an off-center ion and doping with Pr
is not accompanied by creation of the oxygen vacancies in TiO$_6$ octahedra.
The weak effect of the Pr concentration on the temperature of the dielectric
anomaly~\cite{JApplPhys.97.104109}
and a gradual change of the soft mode frequency at the temperature of this
anomaly~\cite{PhysRevB.76.224109} indicate that this anomaly is not
associated with the bulk of the crystal. We think that the model of
ferroelectricity in a system of frozen Pr$^{3+}$--$V_{\rm Sr}$ electric dipoles~\cite{PhysRevB.76.224109}
is also unreasonable because it cannot explain the switching of polarization,
and the local electric fields around defects reduce the lattice susceptibility
and therefore can only decrease the Curie temperature.~\cite{FizTverdTela.35.629}
So, none of the models mentioned in Sec.~\ref{sec1} can explain the
appearance of ferroelectricity.

In our opinion, a possible cause of the dielectric anomaly may be the appearance
of the conductivity associated either with the ionization of the Pr donor levels
(Pr$^{3+}$$\to$~Pr$^{4+}$ + e$^-$) or with a thermally-assisted hopping of
electrons between Pr atoms which can exist in two different oxidation states
(Pr$^{3+}$ and Pr$^{4+}$). Indeed, the results of the recent studies of Pr-doped
SrTiO$_3$ using impedance spectroscopy~\cite{JApplPhys.107.094108} were
interpreted using an equivalent circuit consisting of semiconducting
grains and insulating grain boundaries. In this case, the dispersion of the
dielectric properties is a result of the Arrenius equation for conductivity
of grains.

\section{Conclusions}

X-ray diffraction studies confirmed that the structure of
(Sr$_{1-x}$Pr$_x$)TiO$_3$ solid solutions at 300~K changes from the cubic $Pm3m$
to the tetragonal $I4/mcm$ with increasing $x$. The analysis of the XANES spectra revealed
that the Pr ions are predominantly in the 3+ oxidation state. The analysis of the EXAFS data
showed that the Pr atoms substitute for the Sr atoms in SrTiO$_3$ and are on-center
regardless of the preparation conditions. The local environment of the Pr impurity
(the Pr--O distance) is characterized by a strong relaxation
($\Delta R \approx -0.136$~{\AA}) which is close to the difference of ionic
radii of Sr$^{2+}$ and Pr$^{3+}$. In contrast, in the second shell, a
slight increase in the interatomic distance ($\Delta R \approx +0.02$~{\AA})
was revealed and explained by the repulsion of positively charged Pr$^{3+}$
and Ti$^{4+}$ ions. The weak dependence of the lattice parameter in
(Sr$_{1-x}$Pr$_x$)TiO$_3$ on $x$ is therefore a result of the competition between
the relaxation of the first shell and the repulsion in the second shell.
Various possible defects providing electrical neutrality
of the samples were analyzed and it was shown that the most important
defects in Pr-doped SrTiO$_3$ are charged Sr vacancies and SrO planar faults.
It was shown that among two possible Sr positions in this defect model (Sr in the SrTiO$_3$
lamella-type layers and Sr in the SrO planar faults), the Pr atoms predominantly incorporate
into the SrTiO$_3$ layers.

\begin{acknowledgments}
The authors would like to thank V. F. Kozlovskii for help with x-ray
measurements. I.A.S. and A.I.L. are grateful to Russian-German laboratory
for hospitality and financial support during their stay at BESSY.
\end{acknowledgments}

\bibliographystyle{aipnum4-1}

\providecommand{\BIBYu}{Yu}

\end{document}